\documentclass[a4paper,fleqn]{cas-dc}

\usepackage[authoryear]{natbib}
\usepackage{float}
\usepackage{caption}
\usepackage{booktabs}
\usepackage{tabularx}
\usepackage{stfloats}  % For better placement of floats in two-column format
\usepackage{placeins}
\def\tsc#1{\csdef{#1}{\textsc{\lowercase{#1}}\xspace}}
\tsc{WGM}
\tsc{QE}
\begin{document}
\let\WriteBookmarks\relax
\def\floatpagepagefraction{1}
\def\textpagefraction{.001}

\shorttitle{A combined Machine Learning and Finite Element Modelling tool for the surgical planning of craniosynostosis correction}    

\shortauthors{I.~Antúnez~Sáenz~et~al.}  

\title [mode = title]{A combined Machine Learning and Finite Element Modelling tool for the surgical planning of craniosynostosis correction}  
\author[1,2]{Itxasne {Antúnez Sáenz}}[orcid=0009-0008-4575-2579]
\author[2]{Ane {Alberdi Aramendi}}
\author[1,3]{David Dunaway}
\author[1,3]{Juling Ong}
\author[1]{Lara Deliège}
\author[3]{Amparo Sáenz}
\author[3]{Anita {Ahmadi  Birjandi}}
\author[1,3]{Noor UI Owase Jeelani}
\author[1,3]{Silvia Schievano}
\author[1,4]{Alessandro Borghi}
\cormark[1]
\ead{alessandro.borghi@durham.ac.uk}
% % Corresponding author indication
% \cormark[1]

% % Footnote of the first author

% % Email id of the first author
% \ead{}

% % URL of the first author


\affiliation[1]{organization={Great Ormond Street Institute of Child Health, UCL},
                addressline={30 Guilford Street}, 
                city={London},
                postcode={WC1N 1EH}, 
                country={UK}}

\affiliation[2]{organization={Biomedical Engineering Department, Mondragon Unibertsitatea},
                addressline={Loramendi Kalea 4}, 
                postcode={20500}, 
                %postcodesep={}, 
                city={Arrasate-Mondragón},
                country={Spain}}

\affiliation[3]{organization={Great Ormond Street Hospital},
                addressline={Great Ormond Street}, 
                city={London},
                postcode={WC1N 3JH}, 
                % state={Orissa}, 
                country={UK}}
\affiliation[4]{organization={Department of Engineering, Durham University},
                addressline={Stockton Rd}, 
                city={Durham},
                postcode={DH1 3LE}, 
                % state={Orissa}, 
                country={UK}}
    \cortext[cor1]{Corresponding author}


\begin{abstract}
Craniosynostosis is a medical condition that affects the growth of babies’ heads, caused by an early fusion of cranial sutures. In recent decades, surgical treatments for craniosynostosis have significantly improved, leading to reduced invasiveness, faster recovery, and less blood loss. At Great Ormond Street Hospital (GOSH), the main surgical treatment for patients diagnosed with sagittal craniosynostosis (SC) is spring assisted cranioplasty (SAC). This procedure involves a $15\times15 \text{mm}^2$ osteotomy, where two springs are inserted to induce distraction.\

Despite the numerous advantages of this surgical technique for patients, the outcome remains unpredictable due to the lack of efficient preoperative planning tools. The surgeon’s experience and the baby’s
age are currently relied upon to determine the osteotomy location and spring selection. Previous tools for predicting the surgical outcome of SC relied on finite element modeling (FEM), which involved computed
tomography (CT) imaging and required engineering expertise and lengthy calculations.\

The main goal of this research is to develop a real-time prediction tool for the surgical outcome of patients, eliminating the need for CT scans to minimise radiation exposure during preoperative planning. The proposed methodology involves creating personalised synthetic skulls based on three-dimensional (3D) photographs, incorporating population average values of suture location, skull thickness, and soft tissue properties. A machine learning (ML) surrogate model is employed to achieve the desired surgical outcome.

The resulting multi-output support vector regressor model achieves a $\text{R}^2$ metric of 0.95 and MSE and MAE below 0.13. Furthermore, in the future, this model could not only simulate various surgical scenarios but also provide optimal parameters for achieving a maximum cranial index (CI).
\end{abstract}

%\nocite{*}

% Keywords

\maketitle

% Main text
\section{Introduction}\label{introduction}
Craniosynostosis is a condition that affects approximately 1 in 2000 babies in the UK \citep{Johnson2011PRACTICALGENETICS}. Babies with craniosynostosis not only have an abnormal head shape but may also experience impaired brain growth. In most cases, surgical intervention is required to correct the shape and facilitate proper growth \citep{Morris2016NonsyndromicDisorders}.

Craniosynostosis can be classified into two groups: syndromic craniosynostosis (associated with a genetic abnormality) , and non-syndromic craniosynostosis which represents 92.21 \% of all cases \citep{Tarnow2022IncidenceSweden}.

Craniosynostosis can also be classified according to the affected suture. The most frequent one is sagittal craniosynostosis (with a male-to-female ratio of 3.1 to 1 \citep{Morris2016NonsyndromicDisorders}) where the head midline suture (sagittal suture) is involved. The early fusion of the sagittal suture produces a long and narrow head (scaphocephaly) \citep{Morris2016NonsyndromicDisorders}.

The diagnosis is commonly done by means of a radiologic examination. Following  primary care physician diagnosis, a plain skull radiographic series is ordered to confirm the sutural fusion. Afterwards, computed tomography (CT) imaging is recommended to obtain more information \citep{Fearon2014Evidence-BasedMedicine}.

Even if CT is currently considered the main diagnostic tool for craniosynostosis, it has been demonstrated that 3D-CT images are better than plain radiographs or standard CT \citep{Fearon2014Evidence-BasedMedicine}. The alternatives for avoiding radiation are detecting the suture fusion employing ultrasound technology, whose reliability is still debated \citep{Fearon2014Evidence-BasedMedicine}, or the physical examination  (accurate in 98\% of the cases).

Three-dimensional (3D) photography is a non-invasive and non-ionising method used as an alternative to other 3D imaging types to acquire surface information of the human body \citep{Tenhagen2016Three-DimensionalCraniosynostosis}, \citep{Beaumont2017Three-dimensionalShape}. This technology has been successfully employed in cranial and maxillofacial surgery to compute and quantify changes after surgeries \citep{Tenhagen2016Three-DimensionalCraniosynostosis}, \citep{Rodriguez-Florez2017QuantifyingModelling}.

\begin{figure}[!ht]
    \centering
    \includegraphics[width=\linewidth]{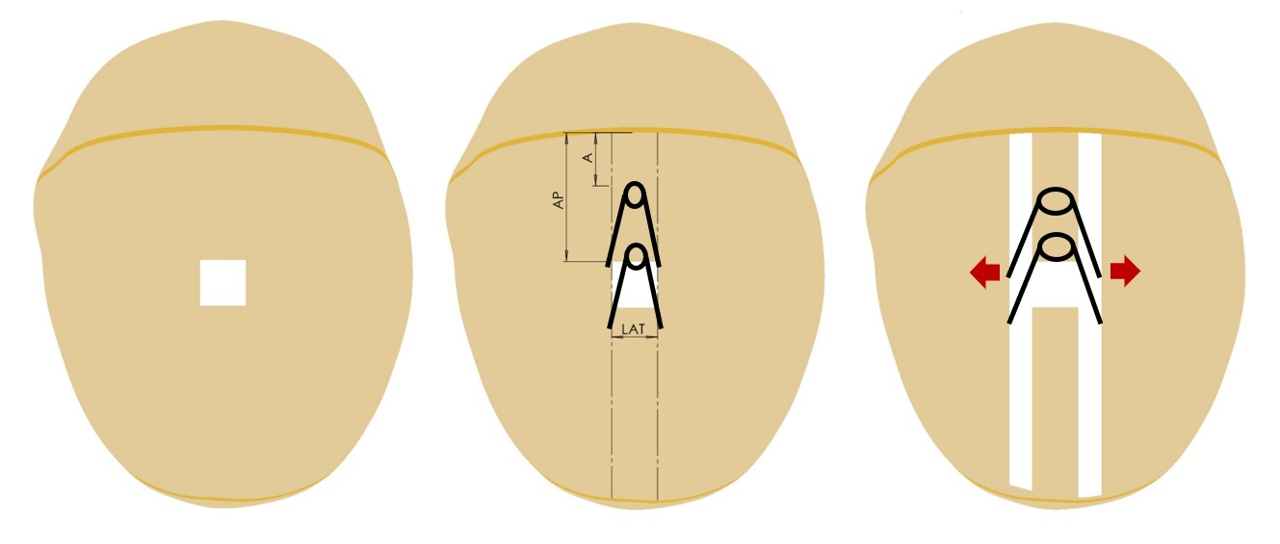}
    \caption{Sketch of the SAC procedure.}
    \label{fig:SAC}
\end{figure}

There are several ways in which craniosynostosis can be treated. Invasive techniques include total calvarial remodelling (TCR) and Distraction for Fronto-orbital or Posterior Cranial Surgery. Both techniques consist on trimming and cutting different parts of the skull to form normal shapes. Nevertheless, also non-invasive techniques are available for treatment. Non-invasive techniques include the Endoscopic strip craniectomy followed by helmet therapy which consists of making small incisions to remove the fused suture. The patient must wear a helmet during the following months after surgery to continue with the modelling. Another non-invasive technique, that currently is the most used one to treat SC at GOSH is spring assisted cranioplasty (SAC). The procedure starts with an incision between 80 and 100 mm over the top of the head that will expose the skull. Once the skull is exposed, an osteotomy will be performed to remove a portion of the fused sagittal sutures and the springs distractors placed as displayed in Figure \ref{fig:SAC}.

\begin{figure}[!ht]
    \centering
    \includegraphics[width=\textwidth, height=5cm]{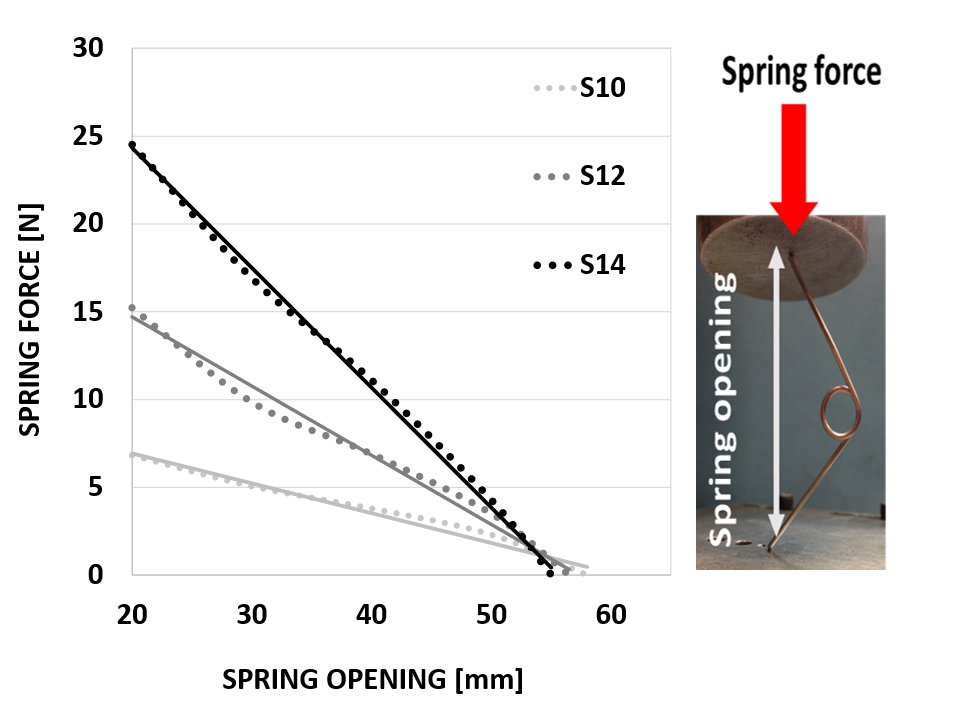}
    \caption{Force vs Opening curve followed by the springs in the uncrimping phase.}
    \label{fig:spring}
\end{figure}

The springs used at GOSH to correct SC, are stainless steel wires with a central loop and an initial opening of 60 mm \citep{Jeelani2020TheSurgery}, \citep{Borghi2017AssessmentPatients}. The springs are crimped before insertion and then they passively expand. During the uncrimping phase the springs follow a Hookean behaviour exerting an outward force directly proportional to the amount of compression they have undergone as displayed in Figure \ref{fig:spring}. 

In previous surgical planning processes, finite element modelling (FEM) has been used to obtain reasonably accurate results. However, this approach requires the use of CT imaging, along with engineering expertise and lengthy calculation times \citep{Borghi2017AssessmentPatients}, \citep{Borghi2018SpringModel}, \citep{GarateAndikoetxea2023TowardsScaphocephaly}. Therefore, to date the surgeon selects the surgical parameters, spring model, and location in the operating theatre based on their own experience and the age of the baby resulting in  unpredictable outcomes \citep{Borghi2017AssessmentPatients}. Thus, a more efficient tool is required to select the appropriate surgical parameters, spring model and location to achieve the best possible result.

Recently, AI-FEM surrogate models (based on machine learning (ML) and Deep Learning (DL)) have been developed for preoperative planning of surgical procedures - such as in the cardiovascular field \citep{Madani2019BridgingAtherosclerosis}. Surrogate models are hybrid models trained using a wide range of FEM simulations, enabling real-time predictions of stresses and deformations without extensive calculations \citep{Karolius2016Multi-scaleSurrogate-models}. 

Statistical shape modelling (SSM) is a dimensionality reduction technique used to describe anatomical variations within a subject population. It has been used in the past for organ, tumour or bone visualization, surgical planning or quantification of disease progression \citep{Iyer2023StatisticalBoundaries} . SSM is based on the quantification of the shape variation after Principal Component Analysis (PCA) is applied in a set of shape vectors \citep{Mei2008StatisticalRetained}. PCA will determine the main components to describe the shape variation within the population. The components that describe the shape variation of the first set of samples can be optimised to fit new individuals, creating a familiar active shape model (ASM) \citep{Mei2008StatisticalRetained}. This method has already been applied in craniofacial surgery \citep{Rodriguez-Florez2017StatisticalCranioplasty}, \citep{Heutinck2021StatisticalVariations}.

Considering the favourable outcomes achieved by FEM in predicting the surgical outcome of craniosynostosis, we hypothesise that an AI-FEM surrogate model could be potentially employed to maintain the positive results offered by FEM for this application, while eliminating the associated challenges, namely CT imaging, engineering expertise, and time consumption.
\newpage
\section{Methods}\label{methods}
\subsection{Data Collection}
\begin{figure}[!ht]
    \centering
    \includegraphics[width = \linewidth]{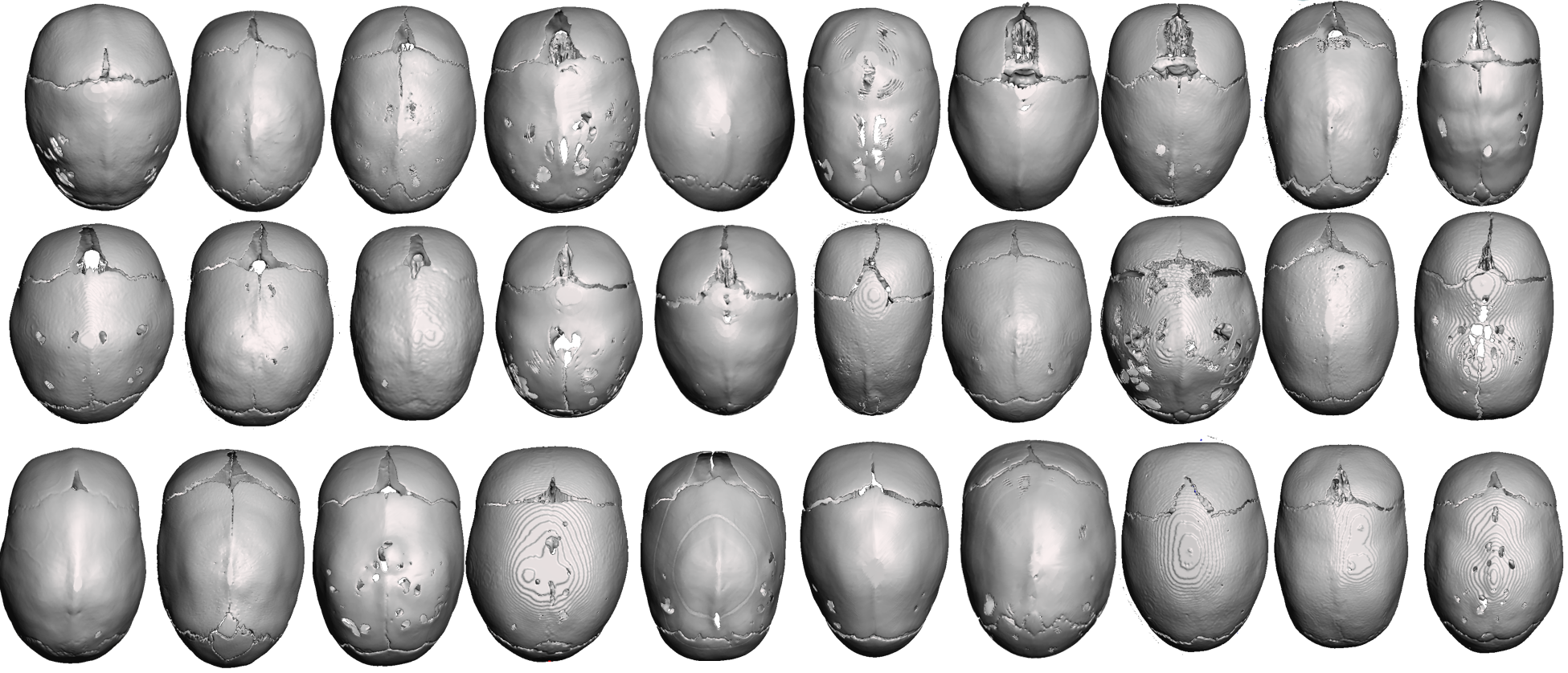}
    \caption{Upper view of the CT scans of the population.}
    \label{fig:CT_pop}
\end{figure}

\begin{figure}[!ht]
    \centering
    \includegraphics[width =  \linewidth]{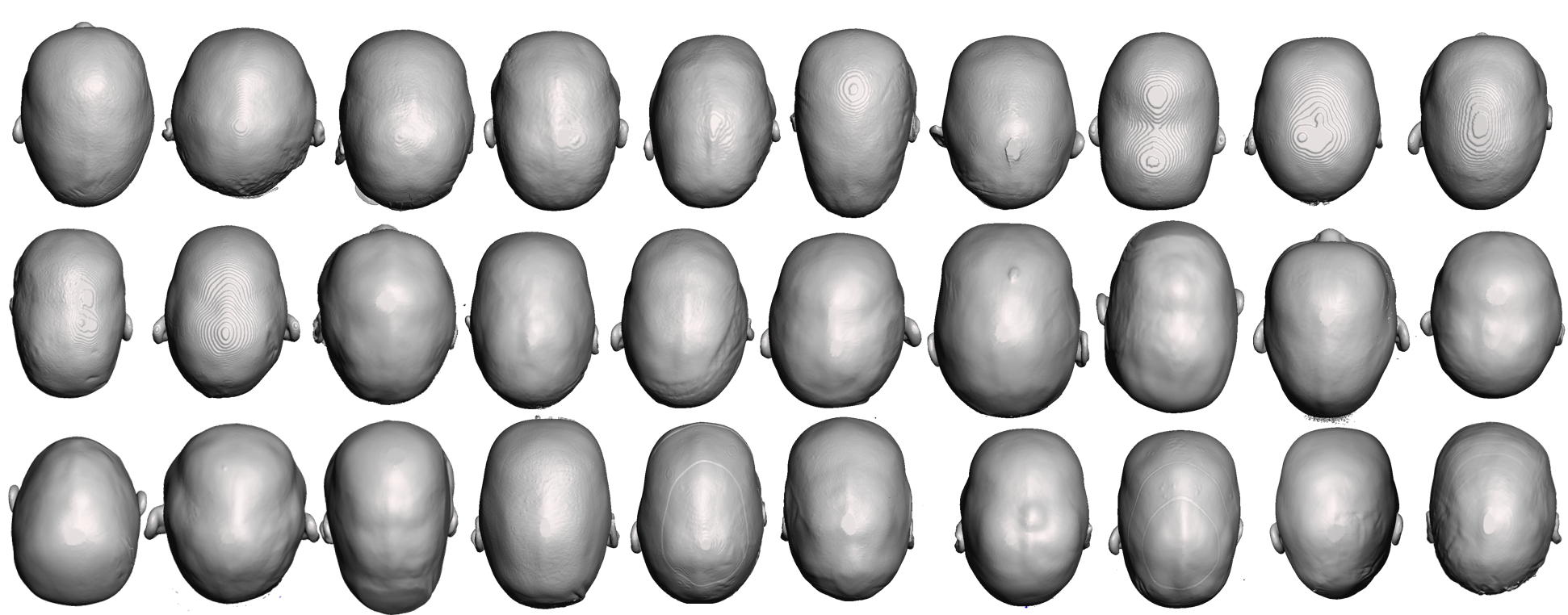}
    \caption{Upper view of the 3D scans of the population.}
    \label{fig:3D_pop}
\end{figure}

In this retrospective study, CT scans were processed to extract 3D skull shapes  (Figure \ref{fig:CT_pop}) and 3D head shapes (Figure \ref{fig:3D_pop}) of 30 non-syndromic sagittal craniosynostosis patients aged $5.8\pm{1.15}$ months at the time of surgery which were intervened at GOSH between December 2011 and June 2022 were employed.

The CT images provided valuable population data on suture location and positioning, as well as skull and soft-tissue thickness.

Additionally, a second separate population consisting of 13 non-syndromic sagittal craniosynostosis patients aged $5.1\pm1.0$ with the preoperative and follow-up (3 weeks after surgery) 3D photographs operated at GOSH with SAC procedure between 2015 and 2017 was used for validation.
\begin{figure}[!ht]
    \centering
    \includegraphics[width =  \linewidth]{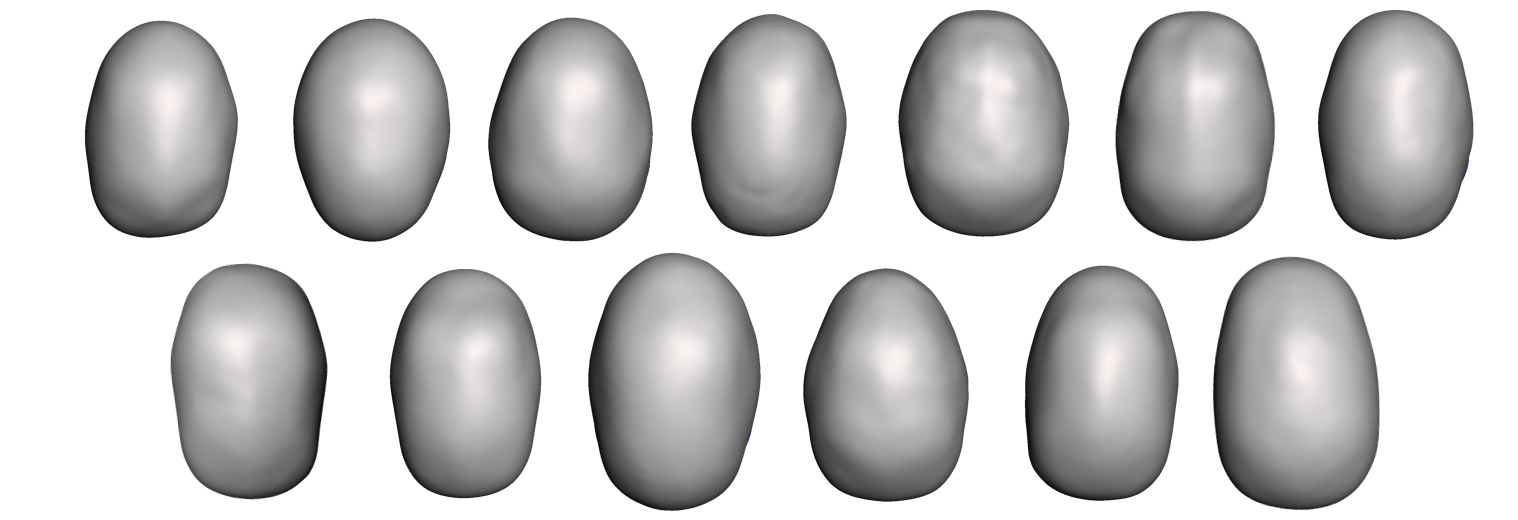}
    \caption{Upper view of the 3D scans of the validation population.}
    \label{fig:val_pop}
\end{figure}
\subsection{Image Processing}
\subsubsection{CT Image Processing}
Head and skull models were extracted in STL format (standard triangulation language), with a 70\% reduction in the number of triangles to mitigate computational costs in subsequent stages. Artifacts in CT images, stemming from both sensor limitations and areas devoid of bone, are common and were covered using Autodesk MeshMixer.

Each skull model was cut along the  nasion-upper auditory meatus plane, which  effectively delineates the neurocranium from the viscerocranium. Anatomical measurements were taken, including the length of the skull, distance between sutures, and angles (following the protocol described in \citep{GarateAndikoetxea2023TowardsScaphocephaly}). All the distances were converted to ratios to ensure robustness.  Each model was split into outer and inner skull to determine individual skull thickness (for this purpose, the lower 25\% was excluded to mitigate potential errors introduced by spurious anatomical structures). Average population skull thickness ($t_{skull}$) was calculated. 

Each head model  was processed similarly: triangle reduction was followed by artifact reduction (performed in Autodesk Meshmixer). Each model was cut along the nasion-tragion plane. Soft tissue thickness was computed using surface distance between outer skull and head model for each patient,  following the protocol described in \citep{GarateAndikoetxea2023TowardsScaphocephaly}). Average population skin thickness ($t_{skin}$) was calculated. 

\subsubsection{3D Image Processing}

Each processed head model was imported into Solidworks \textsuperscript{\textcopyright} where a NURBS was generated ; subsequently, it was uniformly internally offset by $t_{skin}$ (i.e. the population mean of soft tissue thickness). This process yielded a synthetic skull. Sutures were artificially created on the model following the method by \citep{GarateAndikoetxea2023TowardsScaphocephaly} , assuming a width of 2 mm  \citep{Erasmie1976NormalAssessment}.  The final synthetic skull model was composed of frontal bone, parietal bone, occipital bone, coronal and lambdoid suture as displayed in Figure \ref{fig:SAC}. 

Ultimately, the osteotomy was replicated on the parietal bone, following the same method previously published by our group (\citep{Borghi2018SpringModel},\citep{GarateAndikoetxea2023TowardsScaphocephaly}). Briefly, a central point was established at the center of the coronal suture, serving as the origin for all measurements. Subsequently, a rectangle (spanning from the coronal suture to the lambdoid suture) was outlined and cut out of the parietal bone, including four notches (having a 5 mm diameter) replicating the surgical point of insertion of the cranioplasty springs \citep{Rodgers2017Spring-AssistedCases}.

Osteotomies were defined using three measurements: First, the distance between the reference point and the closest notches (anterior spring location - $A$), the distance between the reference point and the other notches $(AP)$ and finally the width of the rectangle $(LAT)$ as displayed on Figure \ref{fig:SAC}.

\subsection{Finite Element Modelling}

The assemblies were imported into Ansys Benchmark \textsuperscript{\textcopyright}. Edges where forces and constraints were applied had to be selected and grouped using named selections. This selection encompassed the fixed bottom edges, the four individual notches where the springs were inserted, and the borders of the five distinct parts that were bonded together.

The material properties assigned to each part were defined based on \citep{Borghi2020AOutcomes}. Frontal, parietal and occipital bones  were assigned isotropic elastic material properties (Young Modulus and Poisson’s ratio) and viscoelastic properties (described through a Prony series).

The spring action was replicated using linear spring conditions, described in terms of stiffness and free length,  with both parameters being parametrised to facilitate adjustments during the design of experiments. The mechanical properties of the three different spring models were obtained from \citep{Borghi2017AssessmentPatients}.

To prevent unrealistic behaviour, bond restrictions were enforced between all parts. In terms of meshing, a standard size of $1.5$ mm was applied to bone, and $1$ mm to sutures. The preferred mesh method is triangulation, and all parts have a consistent width of $t_{skull}$ mm. 

The design of experiments (DoE) was used to generate different surgical configurations and hence increase the dimensionality and variability of the population. The scenarios were automatically created using the optimal-space-filling configuration of Ansys Benchmark \textsuperscript{\textcopyright}.

During the DoE stage, constrains were applied to each parameter. For springs, the range was determined based on the real springs range, while osteotomy parameters were limited according to the specifications outlined in Table \ref{tab:osteotomy_parameters}.

\begin{center}
\captionof{table}{Parameters and ranges.}
\label{tab:osteotomy_parameters}
\begin{tabularx}{0.45\textwidth}{lXlX}
    \toprule
    & \textbf{Parameter}  & \textbf{Value (\%)} \\
    \midrule
    & \textbf{A}  & [18-30] \\
    & \textbf{AP}  & [47-63] \\
    & \textbf{LAT} & [10-25] \\
    \bottomrule
\end{tabularx}
\end{center}

The DoE assigned different values for the different parameters. Surgical parameters followed a normal distribution and the springs models a uniform one.
After configuring all parameters, DoE was conducted for each of the 30 patients, encompassing $80 \pm{5}$ distinct surgical configurations for each. This yielded a total of 2356 FEM outputs. All of these outputs were automatically exported in Compact Database (cdb) format.

\subsection{Data Acquisition}
The data mining process started with the conversion of the cdb files into actual meshes. This process was automated, involving the identification of node and element locations within these files. Then, meshes underwent an external offset of $t_{skin}$ to recover the soft-tissue shape and were saved as STLs. 

The osteotomy gap was closed using Materialise 3-Matic \textsuperscript{\textcopyright}, which maintained the natural roundness of the skull. To achieve point to point correspondence, a template was fitted to each output mesh using Non-rigid Iterative Closest Point (NRICP) algorithm. This template standardizes the number of nodes and elements constituting each of the skulls, which was crucial for the correct implementation of Statistical Shape Modeling (SSM).

\subsubsection{Statistical Shape Modeling (SSM)}
SSM is a technique to describe shapes based on a mean shape \textbf{\textit{$\overline{\mbox{M}}$}} with certain number of modes that represent the variations \textbf{$\Phi$\textit{b}} to obtain the new shape \textbf{\textit{$M_p$}} as shown in Equation \ref{eq1} \citep{Pascoletti2008OFeatures}.

\begin{equation}
    M_p = \overline{M} +\Phi b
    \label{eq1}
\end{equation}

There are 30 shapes corresponding to pre-operatory stage and 2356 from simulated surgical outcomes. The SSM was separately conducted on these two groups with $N_{preop} = 30$ and $N_{postop} = 2356$. Following the procedure described in \citep{Pascoletti2008OFeatures} the two mean shapes of input and outputs sets were achieved \textbf{\textit{$\overline{\mbox{M}}$}} $\in \mathbb{R}^{N \times 3}$.\\

The variability was computed for each of the sets by rearranging the nodes and elements of each i-th shape into column vectors.\\ 
\begin{equation*}
    \textbf{\textit{$M_{preop}$}} \in \mathbb{R}^{N_{preop} \times 3}
\end{equation*}
 \begin{equation*}
     \textbf{\textit{$M_{postop}$}} \in \mathbb{R}^{N_{postop} \times 3}
 \end{equation*}
 
Leading to the following expressions for preoperative shapes:\\
\begin{equation*}
    \textit{\textbf{${m}_{preop , i}$}} \in\mathbb{R}^{N_{preop} \times 1}
\end{equation*}
\begin{equation*}
    \textit{\textbf{$\overline{\mbox{m}}_{preop}$}}\in\mathbb{R}^{N_{preop} \times 1}
\end{equation*}

And postoperative shapes:\\
\begin{equation*}
    \textit{\textbf{${m}_{postop , i}$}} \in\mathbb{R}^{N_{postop} \times 1}
\end{equation*}
\begin{equation*}
    \textit{\textbf{$\overline{\mbox{m}}_{postop}$}}\in\mathbb{R}^{N_{postop} \times 1}
\end{equation*}

After rearranging the following two expressions are obtained:
\begin{equation}
    \textit{\textbf{${m}_{preop , i}$}} = [x_1,y_1,z_1,x_2,y_2,z_2,...,x_{N_{preop}},y_{N_{preop}},z_{N_{preop}}]^T
\end{equation}
\begin{equation}
    \textit{\textbf{${m}_{postop , i}$}} = [x_1,y_1,z_1,x_2,y_2,z_2,...,x_{N_{postop}},y_{N_{postop}},z_{N_{postop}}]^T
\end{equation}

The new representation of the shapes set was computed by the deviation between the individual shapes and the corresponding mean as follows:
\begin{equation}
    \textit{\textbf{${m}_{preop , d}$}} =    \textit{\textbf{${m}_{preop , i}$}} - \textit{\textbf{$\overline{\mbox{m}}_{preop}$}}
\end{equation}
\begin{equation}
    \textit{\textbf{${m}_{postop , d}$}} =    \textit{\textbf{${m}_{postop , i}$}} - \textit{\textbf{$\overline{\mbox{m}}_{postop}$}}
\end{equation}

These vectors were rearranged again into two matrices:\\
\begin{equation*}
    \textbf{\textit{$M_{preop, d}$}} = [m_{preop,1},m_{preop,2},...,m_{preop,N_{preop}}]
\end{equation*}
\begin{equation*}
    \textbf{\textit{$M_{postop, d}$}} = [m_{postop,1},m_{postop,2},...,m_{postop,N_{postop}}]
\end{equation*} 
Eigenvectors and eigenvalues were computed from the covariance matrices of the \textbf{\textit{$M_{preop, d}$}} and \textbf{\textit{$M_{postop, d}$}}. The eigenvectors are given by:\\
\begin{equation*}
    \varphi_i \in \mathbb{R}^{N_{preop} \times 1} i \in [1,N_{preop}-1] 
\end{equation*}
\begin{equation*}
    \varphi_i \in \mathbb{R}^{N_{postop} \times 1} i \in [1,N_{postop}-1]
\end{equation*}

The variances are sorted based on the descriptive variance and corresponding PCs rearranged. Explained variance is given by the following equation:
\begin{equation}
    \lambda_{explained,i} = \frac{\lambda_i}{\sum_{i=1}^{N-1}\lambda_i}
\end{equation}
Finally, SSM was conducted separately on the 30 input shapes and  on the 2356 outcome shapes, which characterized the shapes in terms of a mean shape and variations. To efficiently determine the number of modes required to describe a population, a cumulative distribution function (CDF) was plotted. The modes describing 94\% percentage of variation were selected.

All the collected data was then compiled in a single dataset to train the algorithms. The dataset included the patient's age at surgery time expressed in days, surgical parameters (in ratios), springs' stiffness and free length, and the modes describing both input and output shapes. This compilation resulted in a dataset of size 2356x30.

\subsection{Machine Learning}
The dataset was split into training and testing sets with a test size of 0.33. Seven multi-output machine learning algorithms were evaluated with the main objective of predicting output shapes based on input variables, including age, surgery time, surgical parameters, spring characteristics, and input modes. The tested models included Linear Regression (LR) (used as the baseline), Decision Tree (DT), Random Forest (RF), XGBoost (XGB), Support Vector Machine (SVM), Gradient Boosting (GB), and AdaBoost (AB). 

\begin{figure}[!ht]
    \centering
    \includegraphics[width=\linewidth]{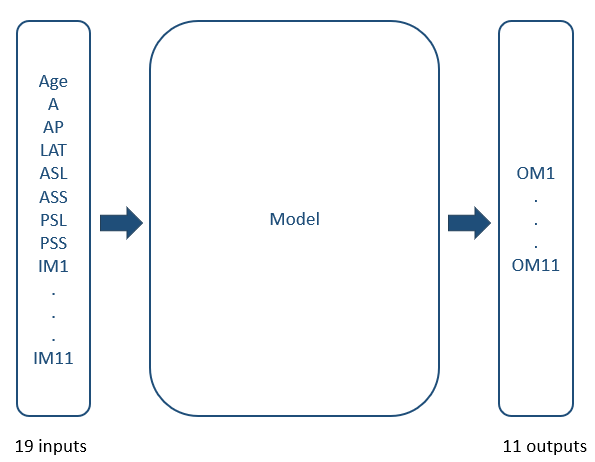}
    \caption{Multi-output regression sketch}
    \label{fig:multi_sketch}
\end{figure}

Since some of these models rely on distance-based calculations, all variables needed to be on the same scale. Therefore, a standard scaler was applied to normalize the input data. The structure of the multi-output regression models is illustrated in Figure \ref{fig:multi_sketch}.

\newpage
Model performance was evaluated using $R^2$, MAE and MSE metrics, along with a 5-fold cross-validation approach. Models exhibiting high variability during 5-fold cross-validation were discarded, as well as those with a lower $R^2$ score than the baseline LR model. Finally, hyperparameter tuning was performed on the four best-performing models using Bayesian search, and their accuracy and error metrics were assessed using the test set.

\subsubsection{Validation using real data}
The validation population was used for the final validation of the selected model. The postoperative 3D photographs were scaled down to omit the growth factor that was not considered at the time of simulation.
\newpage
\section{Results}
\subsection{Image processing}
The skull thickness $t_{skull}$ and the skin thickness $t_{skin}$ of the population were $2.02\pm 0.33$ mm and $3.42\pm0.51$ mm respectively.

\subsection{SSM}
\begin{figure}[!ht]
    \centering
    \includegraphics[width=\linewidth]{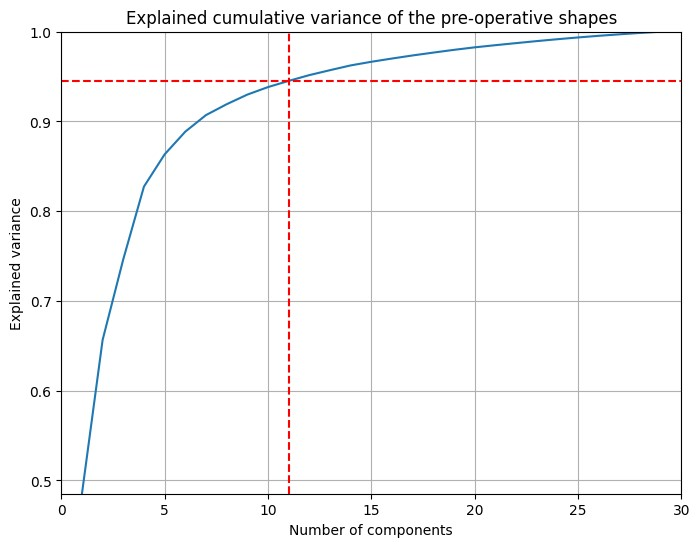}
    \caption{CDF of input modes}
    \label{fig:CDF_in}
\end{figure}

\begin{figure}[!ht]
    \centering
    \includegraphics[width = \linewidth]{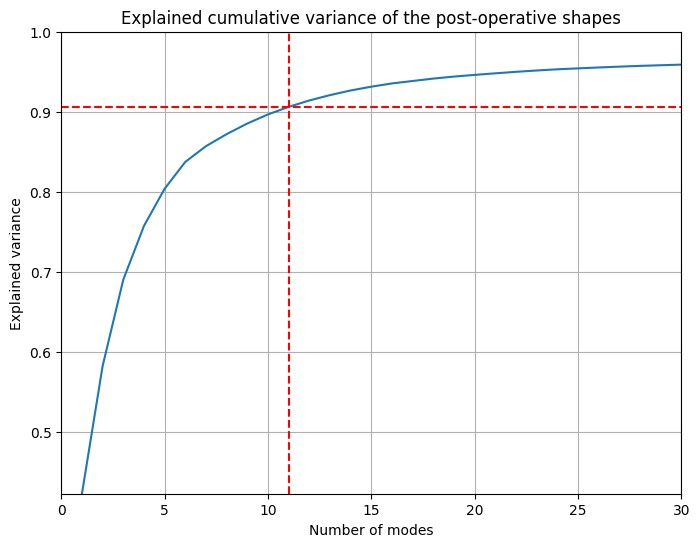}
    \caption{CDF of output modes}
    \label{fig:CDF_out}
\end{figure}

For the input shapes, 11 modes were chosen to comprehensively describe the shapes, accounting for 94\% of the variation. Similarly, the same number of modes (11) were selected for the output, but in this case, they accounted for 90\% of the shape variation as it is illustrated in Figures \ref{fig:CDF_in} and \ref{fig:CDF_out}.

\begin{figure}[!ht]
    \centering
    \includegraphics[width =0.9\linewidth]{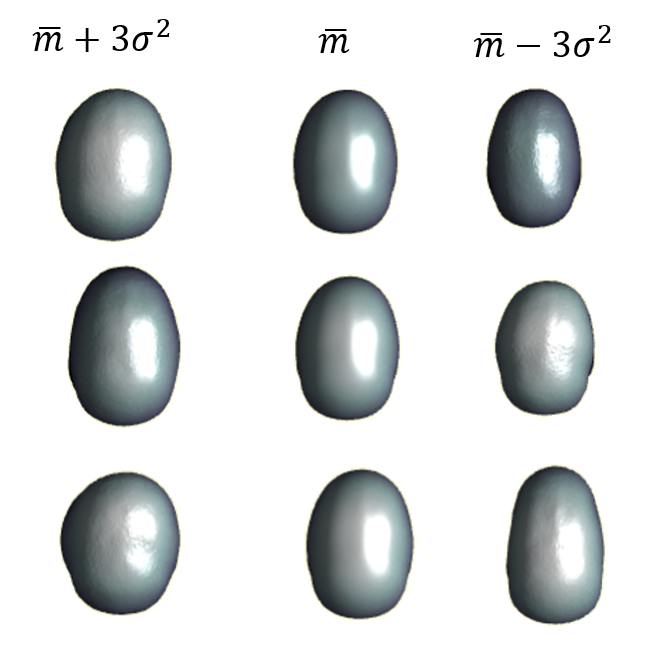}
    \caption{Influence of the input modes.}
    \label{fig:inp_mode}
\end{figure}

\begin{figure}[!ht]
    \centering
    \includegraphics[width = 0.9\linewidth]{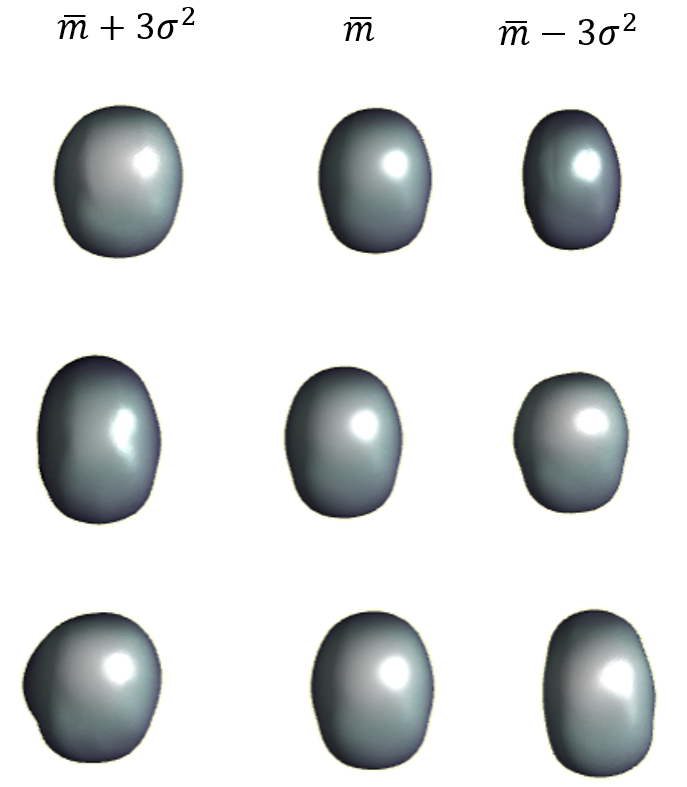}
    \caption{Influence of the output modes.}
    \label{fig:out_mode}
\end{figure}

In the Figures \ref{fig:inp_mode} and \ref{fig:out_mode} the influence of the first 3 modes for input and output can be observed.

\subsection{Predictive Model}
\begin{figure*}[!ht]
    \centering
    \includegraphics[width =\linewidth]{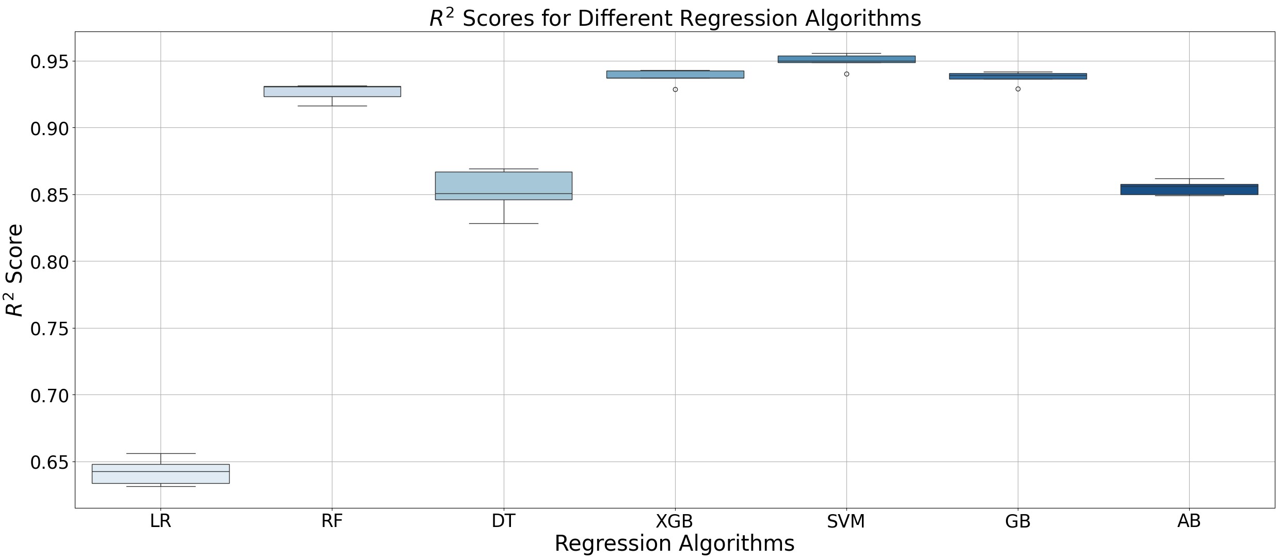}
    \caption{$\text{R}^2$ values scored by the different regression algorithms. Linear Regression (LR), Random Forest (RF), Decision Tree (DT), X Gradient Boosting(XGB), Support Vector Machine (SVM), Gradient Boosting (GB) and AdaBoost (AB).}
    \label{fig:pipeline}
\end{figure*}

In the baseline shown in Figure \ref{fig:pipeline}, the variation of the $\text{R}^2$ value across the 5-folds for different algorithms can be observed. DT exhibited the highest variance, and, taking LR as the baseline, all the remaining models achieved accuracies more than 20 percentage points higher. 

The Table \ref{tab:bayes_parameters} summarizes the search space used for Bayesian hyperparameter optimization for the remaining algorithms, along with the final selected optimal hyperparameter values and their distributions.

\newcolumntype{L}[1]{>{\raggedright\arraybackslash}p{#1}} % Fixed width if needed
\newcolumntype{Y}{>{\raggedright\arraybackslash}X} % Flexible width

\begin{figure*}[!ht]
    \centering
    \includegraphics[width =\linewidth] {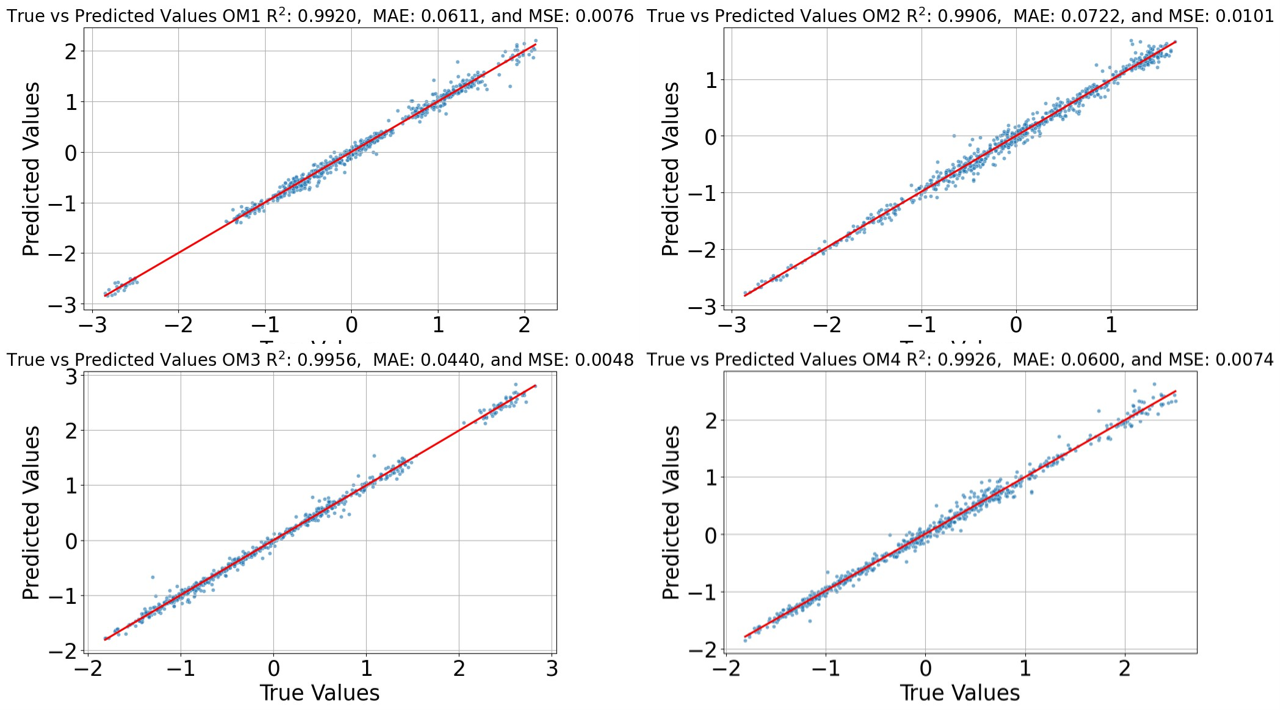}
    \caption{Predicted vs actual values of the first 4 modes.}
    \label{fig:r_2 score}
\end{figure*}

\begin{table*}[!ht]
\centering
\caption{Parameters and distributions used during the Bayesian search optimisation.}
\label{tab:bayes_parameters}
\begin{tabularx}{\textwidth}{llY Y Y c}
    \toprule
    & \textbf{Model} & \textbf{Parameter} & \textbf{Distribution} & \textbf{Value range} & \textbf{Optimal Hyperparameter Value} \\
    \midrule
    \multirow{4}{*}{} & \multirow{4}{*}{\textbf{RF}} 
    & n\_estimators & Uniform & [10-150] & 103 \\
    & & max\_depth & Uniform & [5-20] & 17 \\
    & & min\_samples\_split & Uniform & [2-10] & 2 \\
    & & min\_samples\_leaf & Uniform & [1-5] & 1 \\
    \midrule
    \multirow{5}{*}{} & \multirow{5}{*}{\textbf{XGB}} 
    & booster & Categorical & gbtree, gblinear, dart & dart \\
    & & eta & Uniform & [0.01-1] & 0.18 \\
    & & gamma & Uniform & [0-0.6] & 0.36\\
    & & max\_depth & Uniform & [1-100] & 80 \\
    & & sampling\_method & Categorical & uniform, subsample, gradient\_based & uniform\\
    \midrule
    \multirow{4}{*}{} & \multirow{4}{*}{\textbf{GB}} 
    & n\_estimators & Uniform  & [10-500] & 500 \\
    & & learning\_rate & Uniform & [0.01-1] & 0.12 \\
    & & loss & Categorical & squared\_error, absolute\_error, huber, quantile & huber \\
    & & criterion & Categorical & friedman\_mse, squared\_error & squared\_error \\
    \midrule
    \multirow{5}{*}{} & \multirow{5}{*}{\textbf{SVM}} 
    & degree & Uniform  & [1-9] & 1 \\
    & & gamma & Categorical & scale, auto & auto \\
    & & epsilon & Uniform & [0-5] & 0 \\
    & & kernel & Categorical & linear, poly, rbf, sigmoid & rbf \\
    & & C & Uniform & [0.01-5] & 1.85 \\
    \bottomrule
\end{tabularx}
\end{table*}
\begin{table*}[!ht]
\centering
\caption{$\text{R}^2$, MSE, and MAE of the optimised algorithms.}
\label{tab:bayesian_search}
\begin{tabularx}{0.45\textwidth}{lccc}
    \toprule
    \textbf{Model} & \textbf{$\text{R}^2$} & \textbf{MSE} & \textbf{MAE} \\
    \midrule
    \textbf{RF} & 0.930 & 0.058 & 0.156 \\
    \textbf{XGB} & 0.929 & 0.059 & 0.154 \\
    \textbf{SVM} & 0.950 & 0.042 & 0.122 \\
    \textbf{GB} & 0.939 & 0.050 & 0.137 \\
    \bottomrule
\end{tabularx}
\end{table*}
Final results of the optimised models are shown in Table \ref{tab:bayesian_search}. After analysing these results, SVM emerged as the best-performing algorithm, achieving an $\text{R}^2$ of 0.95 on the test set, with an MAE of 0.122 and an MSE of 0.042. Figure \ref{fig:r_2 score} displays the prediction plot for the different modes.
\subsection{Validation}
Although the algorithms demonstrated accurate results in predicting synthetic data generated by FEM, their performance declined when compared to the actual surgical output. The average metrics of the validation set are: $\text{R}^2$ of -3.58, MSE of 1.41 and MAE of 0.88.
Several factors not included in the FEM, such as growth, may have contributed to the observed differences. 
\newpage
\begin{figure*}[!ht]
    \centering
    \includegraphics[width = 18cm]{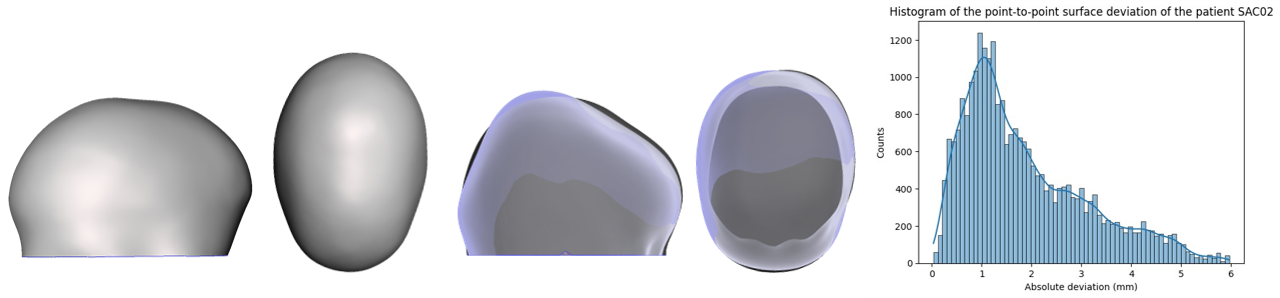}
    \caption{Preoperative 3D Photograph (left), follow-up vs prediction (middle) and distribution of the surface error (right) of a patient with an average error of 1.88 mm.}
    \label{fig:comp}
\end{figure*}

\begin{figure*}[!ht]
    \centering
    \includegraphics[width = 18cm]{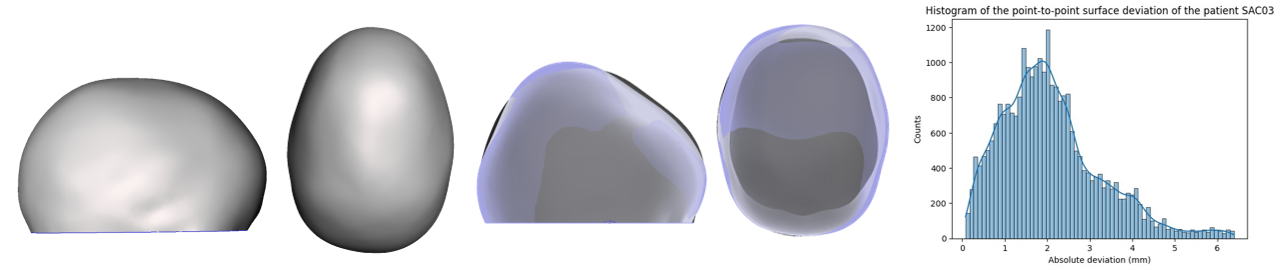}
    \caption{Preoperative 3D Photograph (left), follow-up vs prediction (middle) and distribution of the surface error (right) of a patient with an average error of 2.07 mm.}
    \label{fig:comp1}
\end{figure*}

\begin{figure*}[!ht]
    \centering
    \includegraphics[width = 18cm]{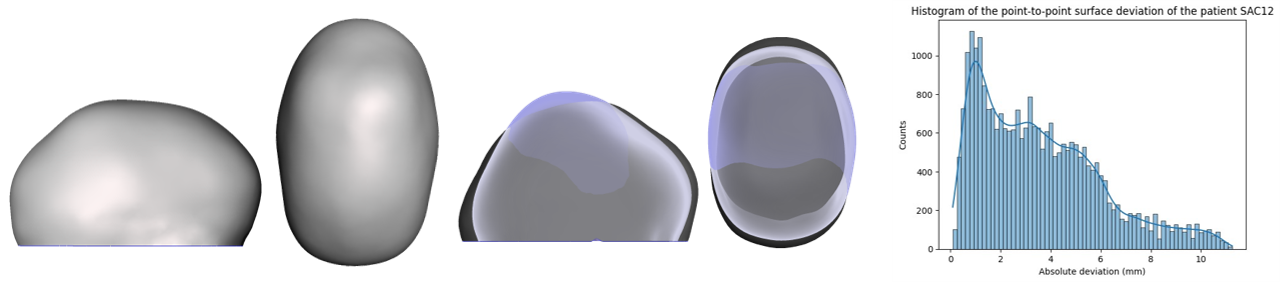}
    \caption{Preoperative 3D Photograph (left), follow-up vs prediction (middle) and distribution of the surface error (right) of a patient with an average error of 3.62 mm.}
    \label{fig:comp2}
\end{figure*}

\begin{figure*}[!ht]
    \centering
    \includegraphics[width = 18cm]{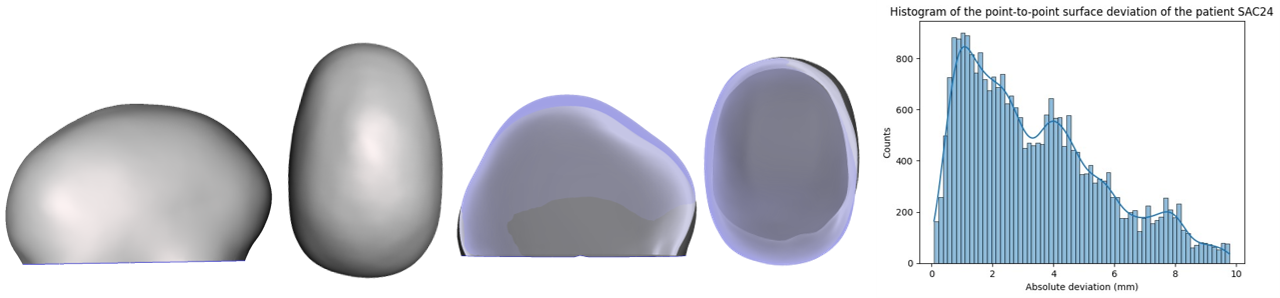}
    \caption{Preoperative 3D Photograph (left), follow-up vs prediction (middle) and distribution of the surface error (right) of a patient with an average error of 3.39 mm.}
    \label{fig:comp3}
\end{figure*}
\clearpage
\section{Discussion and Conclusions}
This paper presents a novel ML-FEM approach to enhance pre-operative planning and outcome predictability for patients undergoing SAC. By leveraging simplified skull models derived from population means, this method facilitates the future use of radiation-free 3D photographs instead of CT scans for preoperative planning. Currently, only 25\% of patients at GOSH receive preoperative imaging to mitigate the risk of tumours later in life \citep{Pearce2012RadiationStudy}.

The feasibility of FEM for predicting surgical outcomes in SAC has been established, validated, and optimized by the research group \citep{Borghi2017AssessmentPatients, Borghi2020AOutcomes, Borghi2018SpringModel, GarateAndikoetxea2023TowardsScaphocephaly}. While traditional numerical models depend on CT imaging data, this study introduces a novel approach that incorporates 3D photographs, thereby eliminating the need for radiation and anaesthesia in young patients. The manual set-up for FEM requires approximately one hour for NURBS creation, and up to 5 hours of simulation to achieve follow-up results for a single surgical configuration. 

Surrogate models trained on FEM data have demonstrated success in other medical fields, such as cardiology \citep{Madani2019BridgingAtherosclerosis}, showcasing their potential for real-time predictions. The developed ML model achieves an $\text{R}^2$ accuracy of 0.95, with MSE and MAE values below 0.13. Despite being trained on synthetic data, it performs real-time predictions with an error rate of $2.70 \pm{0.58}$ mm. This accuracy is comparable to that reported by \citep{Borghi2020AOutcomes}, where 80\% of the points had an error below 2 mm, while also speeding the process and eliminating the need for radiation.

Several assumptions contribute to the model's final errors. The training dataset consisted of a relatively small set of head CT images, while validation was conducted using 3D photographs. Additionally, the model does not account for patient growth between surgery and follow-up, necessitating re-scaling under the assumption of unchanged volume. Other limitations include the use of standardized material properties across the cohort, the absence of suture tearing considerations, and the uniform application of obtained $t_{skull}$ and $t_{skin}$ values to all patients.

Dimensionality reduction techniques should also be further analysed. PCA is a linear dimensionality reduction technique, when not all modes are maintained some explained variance is lost which also contributes to the model error. However, keeping high amount of modes to maintain all the explained variance and hence increasing the dimensionality of the dataset compromises the so called curse of dimensionality introduced by \citep{Bellman1957DYNAMICPROGRAMMING}.  

Future work aims to address these assumptions and thereby reduce errors. Models capable of inferring skull structures from 3D photographs \citep{Milojevic2024AutoSkull:Pipelines} will enable the application of the already validated numerical models to more sophisticated synthetic skulls, moving beyond uniform offsets or population means. 

The synthetic 80 scenarios generated during the DoE stage for each patient introduced variability into the model, allowing it to learn different possible surgical configurations and their effects on the final outcome. Training on real patient data might have improved validation accuracy by reducing modelling errors. However, the limited number of patients and the similarity of surgical configurations would have resulted in a smaller dataset with lower variability.

Currently, the model is capable of real-time predictions of the follow-up results only relying in 3D photographs, in some iterations it will offer clinicians valuable guidance during SAC pre-operative planning. The use of surrogate models has significantly reduced computational time while maintaining predictive accuracy, enabling surgeons to make informed decisions efficiently. This advancement not only enhances preoperative planning but also improves patient and family experience by providing a reliable visual representation of expected surgical outcomes.

Nevertheless, it is essential to acknowledge the study’s limitations. The model was trained and tested on FEM output, disregarding factors such as scalp growth and relying on population-averaged values instead of personalized parameters. These simplifications have led to discrepancies between the synthetic data used for ML model training and actual surgical outcomes. Despite this, the ML model can predict surgical outcomes with an error of $2.70 \pm{0.58}$ mm, delivering results comparable to FEM simulations within seconds.

In conclusion, despite its current limitations, this study underscores the potential of ML-based models in predicting surgical outcomes. Future iterations incorporating real patient data and more realistic simulations are expected to enhance accuracy and reliability, further establishing ML as a valuable tool in preoperative surgical planning.
\section*{Acknowledgements}
The work has been funded by Great Ormond Street Hospital for Children Charity (grant number 12SG15) as well as the NIHR Biomedical Research Centre Advanced Therapies for Structural Malformations and Tissue Damage pump-prime funding call (grant n. 17DS18), the Great Ormond Street Hospital Charity Clinical Research Starter Grant (grant n. 17DD46) and the European Research Council (ERC-2017-StG- 757923). This report incorporates independent research from the National Institute for Health Research Biomedical Research Centre Funding Scheme. The views expressed in this publication are those of the author(s) and not necessarily those of the NHS, the National Institute for Health Research or the Department of Health.

\printcredits

\bibliographystyle{cas-model2-names}

\clearpage
\bibliography{references}

\begin{thebibliography}{24}
\expandafter\ifx\csname natexlab\endcsname\relax\def\natexlab#1{#1}\fi
\providecommand{\url}[1]{\texttt{#1}}
\providecommand{\href}[2]{#2}
\providecommand{\path}[1]{#1}
\providecommand{\DOIprefix}{doi:}
\providecommand{\ArXivprefix}{arXiv:}
\providecommand{\URLprefix}{URL: }
\providecommand{\Pubmedprefix}{pmid:}
\providecommand{\doi}[1]{\href{http://dx.doi.org/#1}{\path{#1}}}
\providecommand{\Pubmed}[1]{\href{pmid:#1}{\path{#1}}}
\providecommand{\bibinfo}[2]{#2}
\ifx\xfnm\relax \def\xfnm[#1]{\unskip,\space#1}\fi
%Type = Article
\bibitem[{Beaumont et~al.(2017)Beaumont, Knoops, Borghi, Owase~Jeelani, Koudstaal, Schievano, Dunaway and Rodriguez-Florez}]{Beaumont2017Three-dimensionalShape}
\bibinfo{author}{Beaumont, C.A.A.}, \bibinfo{author}{Knoops, P.G.M.}, \bibinfo{author}{Borghi, A.}, \bibinfo{author}{Owase~Jeelani, N.U.}, \bibinfo{author}{Koudstaal, M.J.}, \bibinfo{author}{Schievano, S.}, \bibinfo{author}{Dunaway, D.J.}, \bibinfo{author}{Rodriguez-Florez, N.}, \bibinfo{year}{2017}.
\newblock \bibinfo{title}{{Three-dimensional surface scanners compared with standard anthropometric measurements for head shape}}.
\newblock \bibinfo{journal}{Journal of Cranio-Maxillo-Facial Surgery} \URLprefix \url{http://dx.doi.org/10.1016/j.jcms.2017.03.003}, \DOIprefix\doi{10.1016/j.jcms.2017.03.003}.
%Type = Book
\bibitem[{Bellman(1957)}]{Bellman1957DYNAMICPROGRAMMING}
\bibinfo{author}{Bellman, R.}, \bibinfo{year}{1957}.
\newblock \bibinfo{title}{{DYNAMIC PROGRAMMING}}.
\newblock \bibinfo{edition}{6} ed.
%Type = Article
\bibitem[{Borghi et~al.(2018)Borghi, Rodriguez-Florez, Rodgers, James, Hayward, Dunaway, Jeelani and Schievano}]{Borghi2018SpringModel}
\bibinfo{author}{Borghi, A.}, \bibinfo{author}{Rodriguez-Florez, N.}, \bibinfo{author}{Rodgers, W.}, \bibinfo{author}{James, G.}, \bibinfo{author}{Hayward, R.}, \bibinfo{author}{Dunaway, D.}, \bibinfo{author}{Jeelani, O.}, \bibinfo{author}{Schievano, S.}, \bibinfo{year}{2018}.
\newblock \bibinfo{title}{{Spring assisted cranioplasty: A patient specific computational model}}.
\newblock \bibinfo{journal}{Medical Engineering {\&} Physics} \bibinfo{volume}{53}, \bibinfo{pages}{58--65}.
\newblock \DOIprefix\doi{10.1016/J.MEDENGPHY.2018.01.001}.
%Type = Article
\bibitem[{Borghi et~al.(2020)Borghi, Rodriguez~Florez, Ruggiero, James, O'hara, Ong, Jeelani, Dunaway and Schievano}]{Borghi2020AOutcomes}
\bibinfo{author}{Borghi, A.}, \bibinfo{author}{Rodriguez~Florez, N.}, \bibinfo{author}{Ruggiero, F.}, \bibinfo{author}{James, G.}, \bibinfo{author}{O'hara, J.}, \bibinfo{author}{Ong, J.}, \bibinfo{author}{Jeelani, O.}, \bibinfo{author}{Dunaway, D.}, \bibinfo{author}{Schievano, S.}, \bibinfo{year}{2020}.
\newblock \bibinfo{title}{{A population-specific material model for sagittal craniosynostosis to predict surgical shape outcomes}}.
\newblock \bibinfo{journal}{Biomechanics and Modeling in Mechanobiology} \bibinfo{volume}{19}, \bibinfo{pages}{1319--1329}.
\newblock \URLprefix \url{https://doi.org/10.1007/s10237-019-01229-y}, \DOIprefix\doi{10.1007/s10237-019-01229-y}.
%Type = Article
\bibitem[{Borghi et~al.(2017)Borghi, Schievano, Florez, McNicholas, Rodgers, Ponniah, James, Hayward, Dunaway and Owase~Jeelani}]{Borghi2017AssessmentPatients}
\bibinfo{author}{Borghi, A.}, \bibinfo{author}{Schievano, S.}, \bibinfo{author}{Florez, N.R.}, \bibinfo{author}{McNicholas, R.}, \bibinfo{author}{Rodgers, W.}, \bibinfo{author}{Ponniah, A.}, \bibinfo{author}{James, G.}, \bibinfo{author}{Hayward, R.}, \bibinfo{author}{Dunaway, D.}, \bibinfo{author}{Owase~Jeelani, N.U.}, \bibinfo{year}{2017}.
\newblock \bibinfo{title}{{Assessment of spring cranioplasty biomechanics in sagittal craniosynostosis patients}}.
\newblock \bibinfo{journal}{Journal of Neurosurgery: Pediatrics} \bibinfo{volume}{20}, \bibinfo{pages}{400--409}.
\newblock \URLprefix \url{https://thejns.org/pediatrics/view/journals/j-neurosurg-pediatr/20/5/article-p400.xml}, \DOIprefix\doi{10.3171/2017.1.PEDS16475}.
%Type = Article
\bibitem[{Erasmie and Ringertz(1976)}]{Erasmie1976NormalAssessment}
\bibinfo{author}{Erasmie, U.}, \bibinfo{author}{Ringertz, H.}, \bibinfo{year}{1976}.
\newblock \bibinfo{title}{{Normal width of cranial sutures in the neonate and infant. An objective method of assessment}}.
\newblock \bibinfo{journal}{Acta Radiologica - Series Diagnosis} \bibinfo{volume}{17}, \bibinfo{pages}{565--572}.
\newblock \URLprefix \url{https://journals.sagepub.com/doi/10.1177/028418517601705A03}, \DOIprefix\doi{10.1177/028418517601705A03/ASSET/028418517601705A03.FP.PNG{\_}V03}.
%Type = Article
\bibitem[{Fearon(2014)}]{Fearon2014Evidence-BasedMedicine}
\bibinfo{author}{Fearon, J.A.}, \bibinfo{year}{2014}.
\newblock \bibinfo{title}{{Evidence-Based Medicine}}.
\newblock \bibinfo{journal}{Plastic and Reconstructive Surgery} \bibinfo{volume}{133}, \bibinfo{pages}{1261--1275}.
\newblock \URLprefix \url{https://journals.lww.com/plasreconsurg/fulltext/2014/05000/evidence_based_medicine__craniosynostosis.36.aspx}, \DOIprefix\doi{10.1097/PRS.0000000000000093}.
%Type = Article
\bibitem[{Garate~Andikoetxea et~al.(2023)Garate~Andikoetxea, Ajami, Rodriguez-Florez, Jeelani, Dunaway, Schievano, Borghi and Owase~Jeelani}]{GarateAndikoetxea2023TowardsScaphocephaly}
\bibinfo{author}{Garate~Andikoetxea, B.}, \bibinfo{author}{Ajami, S.}, \bibinfo{author}{Rodriguez-Florez, N.}, \bibinfo{author}{Jeelani, N.U.O.}, \bibinfo{author}{Dunaway, D.}, \bibinfo{author}{Schievano, S.}, \bibinfo{author}{Borghi, A.}, \bibinfo{author}{Owase~Jeelani, N.U.}, \bibinfo{year}{2023}.
\newblock \bibinfo{title}{{Towards a radiation free numerical modelling framework to predict spring assisted correction of scaphocephaly}}.
\newblock \bibinfo{journal}{Computer Methods in Biomechanics and Biomedical Engineering} \URLprefix \url{https://www.tandfonline.com/action/journalInformation?journalCode=gcmb20}, \DOIprefix\doi{10.1080/10255842.2023.2294262}.
%Type = Article
\bibitem[{Heutinck et~al.(2021)Heutinck, Knoops, Florez, Biffi, Breakey, James, Koudstaal, Schievano, Dunaway, Jeelani and Borghi}]{Heutinck2021StatisticalVariations}
\bibinfo{author}{Heutinck, P.}, \bibinfo{author}{Knoops, P.}, \bibinfo{author}{Florez, N.R.}, \bibinfo{author}{Biffi, B.}, \bibinfo{author}{Breakey, W.}, \bibinfo{author}{James, G.}, \bibinfo{author}{Koudstaal, M.}, \bibinfo{author}{Schievano, S.}, \bibinfo{author}{Dunaway, D.}, \bibinfo{author}{Jeelani, O.}, \bibinfo{author}{Borghi, A.}, \bibinfo{year}{2021}.
\newblock \bibinfo{title}{{Statistical shape modelling for the analysis of head shape variations}}.
\newblock \bibinfo{journal}{Journal of Cranio-Maxillofacial Surgery} \bibinfo{volume}{49}, \bibinfo{pages}{449--455}.
\newblock \DOIprefix\doi{10.1016/J.JCMS.2021.02.020}.
%Type = Article
\bibitem[{Iyer et~al.(2023)Iyer, Morris, Zenger, Karanth, Khan, Orkild, Korshak and Elhabian}]{Iyer2023StatisticalBoundaries}
\bibinfo{author}{Iyer, K.}, \bibinfo{author}{Morris, A.}, \bibinfo{author}{Zenger, B.}, \bibinfo{author}{Karanth, K.}, \bibinfo{author}{Khan, N.}, \bibinfo{author}{Orkild, B.A.}, \bibinfo{author}{Korshak, O.}, \bibinfo{author}{Elhabian, S.}, \bibinfo{year}{2023}.
\newblock \bibinfo{title}{{Statistical shape modeling of multi-organ anatomies with shared boundaries}}.
\newblock \bibinfo{journal}{Frontiers in Bioengineering and Biotechnology} \bibinfo{volume}{10}, \bibinfo{pages}{1078800}.
\newblock \DOIprefix\doi{10.3389/FBIOE.2022.1078800/BIBTEX}.
%Type = Article
\bibitem[{Jeelani et~al.(2020)Jeelani, Borghi, Florez, Bozkurt, Dunaway and Schievano}]{Jeelani2020TheSurgery}
\bibinfo{author}{Jeelani, N.u.O.}, \bibinfo{author}{Borghi, A.}, \bibinfo{author}{Florez, N.R.}, \bibinfo{author}{Bozkurt, S.}, \bibinfo{author}{Dunaway, D.}, \bibinfo{author}{Schievano, S.}, \bibinfo{year}{2020}.
\newblock \bibinfo{title}{{The science behind the springs: Using biomechanics and finite element modeling to predict outcomes in spring-assisted sagittal synostosis surgery}}.
\newblock \bibinfo{journal}{Journal of Craniofacial Surgery} \bibinfo{volume}{31}, \bibinfo{pages}{2074--2078}.
\newblock \URLprefix \url{https://journals.lww.com/jcraniofacialsurgery/fulltext/2020/10000/the_science_behind_the_springs__using_biomechanics.59.aspx}, \DOIprefix\doi{10.1097/SCS.0000000000006865}.
%Type = Article
\bibitem[{Johnson and M~Wilkie(2011)}]{Johnson2011PRACTICALGENETICS}
\bibinfo{author}{Johnson, D.}, \bibinfo{author}{M~Wilkie, A.O.}, \bibinfo{year}{2011}.
\newblock \bibinfo{title}{{PRACTICAL GENETICS}}.
\newblock \bibinfo{journal}{European Journal of Human Genetics} \URLprefix \url{www.nature.com/ejhg}, \DOIprefix\doi{10.1038/ejhg.2010.235}.
%Type = Article
\bibitem[{Karolius et~al.(2016)Karolius, Preisig and Rusche}]{Karolius2016Multi-scaleSurrogate-models}
\bibinfo{author}{Karolius, S.}, \bibinfo{author}{Preisig, H.A.}, \bibinfo{author}{Rusche, H.}, \bibinfo{year}{2016}.
\newblock \bibinfo{title}{{Multi-scale modelling software framework facilitating simulation of interconnected scales using surrogate-models}}.
\newblock \bibinfo{journal}{Computer Aided Chemical Engineering} \bibinfo{volume}{38}, \bibinfo{pages}{463--468}.
\newblock \DOIprefix\doi{10.1016/B978-0-444-63428-3.50082-5}.
%Type = Article
\bibitem[{Madani et~al.(2019)Madani, Bakhaty, Kim, Mubarak and Mofrad}]{Madani2019BridgingAtherosclerosis}
\bibinfo{author}{Madani, A.}, \bibinfo{author}{Bakhaty, A.}, \bibinfo{author}{Kim, J.}, \bibinfo{author}{Mubarak, Y.}, \bibinfo{author}{Mofrad, M.R.K.}, \bibinfo{year}{2019}.
\newblock \bibinfo{title}{{Bridging Finite Element and Machine Learning Modeling: Stress Prediction of Arterial Walls in Atherosclerosis}}.
\newblock \bibinfo{journal}{Journal of biomechanical engineering} \URLprefix \url{http://asmedigitalcollection.asme.org/biomechanical/article-pdf/141/8/084502/6390325/bio_141_08_084502.pdf?casa_token=eXUNg825u2YAAAAA:RszkB4rS544X-4mPfFYYGmYH3UJqpEjFXt3TgOKEaJWdis0o0pvdDtU5oqQVg1uEwlbgIu-c}, \DOIprefix\doi{10.1115/1.4043290}.
%Type = Article
\bibitem[{Mei et~al.(2008)Mei, Figl, Rueckert, Darzi and Edwards}]{Mei2008StatisticalRetained}
\bibinfo{author}{Mei, L.}, \bibinfo{author}{Figl, M.}, \bibinfo{author}{Rueckert, D.}, \bibinfo{author}{Darzi, A.}, \bibinfo{author}{Edwards, P.}, \bibinfo{year}{2008}.
\newblock \bibinfo{title}{{Statistical shape modelling: How many modes should be retained?}}
\newblock \bibinfo{journal}{2008 IEEE Computer Society Conference on Computer Vision and Pattern Recognition Workshops, CVPR Workshops} \DOIprefix\doi{10.1109/CVPRW.2008.4562996}.
%Type = Inproceedings
\bibitem[{Milojevic et~al.(2024)Milojevic, Peter, Huber, Azevedo, Latyshev, Sailer, Gross, Thomaszewski, Solenthaler and G{\"{o}}zc{\"{u}}}]{Milojevic2024AutoSkull:Pipelines}
\bibinfo{author}{Milojevic, A.}, \bibinfo{author}{Peter, D.}, \bibinfo{author}{Huber, N.B.}, \bibinfo{author}{Azevedo, L.}, \bibinfo{author}{Latyshev, A.}, \bibinfo{author}{Sailer, I.}, \bibinfo{author}{Gross, M.}, \bibinfo{author}{Thomaszewski, B.}, \bibinfo{author}{Solenthaler, B.}, \bibinfo{author}{G{\"{o}}zc{\"{u}}, B.}, \bibinfo{year}{2024}.
\newblock \bibinfo{title}{{AutoSkull: Learning-based Skull Estimation for Automated Pipelines}}, in: \bibinfo{booktitle}{MICCAI}, \bibinfo{publisher}{Springer}. pp. \bibinfo{pages}{109--118}.
\newblock \DOIprefix\doi{10.1007/978-3-031-72104-5{\_}11}.
%Type = Article
\bibitem[{Morris(2016)}]{Morris2016NonsyndromicDisorders}
\bibinfo{author}{Morris, L.M.}, \bibinfo{year}{2016}.
\newblock \bibinfo{title}{{Nonsyndromic Craniosynostosis and Deformational Head Shape Disorders}}.
\newblock \bibinfo{journal}{Facial Plastic Surgery Clinics of North America} \bibinfo{volume}{24}, \bibinfo{pages}{517--530}.
\newblock \DOIprefix\doi{10.1016/J.FSC.2016.06.007}.
%Type = Article
\bibitem[{Pascoletti(2008)}]{Pascoletti2008OFeatures}
\bibinfo{author}{Pascoletti, G.}, \bibinfo{year}{2008}.
\newblock \bibinfo{title}{{O R I G I N A L P A P E R Statistical shape modelling of the human mandible: 3D shape predictions based on external morphometric features}}.
\newblock \bibinfo{journal}{International Journal on Interactive Design and Manufacturing (IJIDeM)} \bibinfo{volume}{16}, \bibinfo{pages}{1675--1693}.
\newblock \URLprefix \url{https://doi.org/10.1007/s12008-022-00882-5}, \DOIprefix\doi{10.1007/s12008-022-00882-5}.
%Type = Article
\bibitem[{Pearce et~al.(2012)Pearce, Salotti, Howe, Little, Lee, Ronckers, Rajaraman, Korea, Pearce, Salotti, Little, McHugh, Lee, Pyo~Kim, Howe, Ronckers, Rajaraman, Alan~Craft, Parker and Berrington~de Gonz{\'{a}}lez}]{Pearce2012RadiationStudy}
\bibinfo{author}{Pearce, M.S.}, \bibinfo{author}{Salotti, J.A.}, \bibinfo{author}{Howe, N.L.}, \bibinfo{author}{Little, M.P.}, \bibinfo{author}{Lee, C.}, \bibinfo{author}{Ronckers, C.M.}, \bibinfo{author}{Rajaraman, P.}, \bibinfo{author}{Korea, S.}, \bibinfo{author}{Pearce, M.S.}, \bibinfo{author}{Salotti, J.A.}, \bibinfo{author}{Little, M.P.}, \bibinfo{author}{McHugh, K.}, \bibinfo{author}{Lee, C.}, \bibinfo{author}{Pyo~Kim, K.}, \bibinfo{author}{Howe, N.L.}, \bibinfo{author}{Ronckers, C.M.}, \bibinfo{author}{Rajaraman, P.}, \bibinfo{author}{Alan~Craft, S.W.}, \bibinfo{author}{Parker, L.}, \bibinfo{author}{Berrington~de Gonz{\'{a}}lez, A.}, \bibinfo{year}{2012}.
\newblock \bibinfo{title}{{Radiation exposure from CT scans in childhood and subsequent risk of leukaemia and brain tumours: a retrospective cohort study}}.
\newblock \bibinfo{journal}{The Lancet} \bibinfo{volume}{380}, \bibinfo{pages}{499--505}.
\newblock \URLprefix \url{http://dx.doi.org/10.1016/}, \DOIprefix\doi{10.1016/S0140-6736(12)60815-0}.
%Type = Inproceedings
\bibitem[{Rodgers et~al.(2017)Rodgers, Glass, Schievano, Borghi, Rodriguez-Florez, Tahim, Angullia, Breakey, Knoops, Tenhagen, O'Hara, Ponniah, James, Dunaway and Jeelani}]{Rodgers2017Spring-AssistedCases}
\bibinfo{author}{Rodgers, W.}, \bibinfo{author}{Glass, G.E.}, \bibinfo{author}{Schievano, S.}, \bibinfo{author}{Borghi, A.}, \bibinfo{author}{Rodriguez-Florez, N.}, \bibinfo{author}{Tahim, A.}, \bibinfo{author}{Angullia, F.}, \bibinfo{author}{Breakey, W.}, \bibinfo{author}{Knoops, P.}, \bibinfo{author}{Tenhagen, M.}, \bibinfo{author}{O'Hara, J.}, \bibinfo{author}{Ponniah, A.}, \bibinfo{author}{James, G.}, \bibinfo{author}{Dunaway, D.J.}, \bibinfo{author}{Jeelani, N.U.}, \bibinfo{year}{2017}.
\newblock \bibinfo{title}{{Spring-Assisted Cranioplasty for the Correction of Nonsyndromic Scaphocephaly: A Quantitative Analysis of 100 Consecutive Cases}}, in: \bibinfo{booktitle}{Plastic and Reconstructive Surgery}, \bibinfo{publisher}{Lippincott Williams and Wilkins}. pp. \bibinfo{pages}{125--134}.
\newblock \DOIprefix\doi{10.1097/PRS.0000000000003465}.
%Type = Article
\bibitem[{Rodriguez-Florez et~al.(2017a)Rodriguez-Florez, Bruse, Borghi, Vercruysse, Ong, James, Pennec, Dunaway, Owase~Jeelani and Schievano}]{Rodriguez-Florez2017StatisticalCranioplasty}
\bibinfo{author}{Rodriguez-Florez, N.}, \bibinfo{author}{Bruse, J.L.}, \bibinfo{author}{Borghi, A.}, \bibinfo{author}{Vercruysse, H.}, \bibinfo{author}{Ong, J.}, \bibinfo{author}{James, G.}, \bibinfo{author}{Pennec, X.}, \bibinfo{author}{Dunaway, D.J.}, \bibinfo{author}{Owase~Jeelani, N.U.}, \bibinfo{author}{Schievano, S.}, \bibinfo{year}{2017}a.
\newblock \bibinfo{title}{{Statistical shape modelling to aid surgical planning: associations between surgical parameters and head shapes following spring-assisted cranioplasty}}.
\newblock \bibinfo{journal}{International Journal of Computer Assisted Radiology and Surgery} \bibinfo{volume}{12}, \bibinfo{pages}{1739--1749}.
\newblock \DOIprefix\doi{10.1007/s11548-017-1614-5}.
%Type = Article
\bibitem[{Rodriguez-Florez et~al.(2017b)Rodriguez-Florez, G{\"{o}}ktekin, Bruse, Borghi, Angullia, Knoops, Tenhagen, O~Hara, Koudstaal, Schievano, Owase~Jeelani, James and Dunaway}]{Rodriguez-Florez2017QuantifyingModelling}
\bibinfo{author}{Rodriguez-Florez, N.}, \bibinfo{author}{G{\"{o}}ktekin, O.K.}, \bibinfo{author}{Bruse, J.L.}, \bibinfo{author}{Borghi, A.}, \bibinfo{author}{Angullia, F.}, \bibinfo{author}{Knoops, P.G.M.}, \bibinfo{author}{Tenhagen, M.}, \bibinfo{author}{O~Hara, J.L.}, \bibinfo{author}{Koudstaal, M.J.}, \bibinfo{author}{Schievano, S.}, \bibinfo{author}{Owase~Jeelani, N.U.}, \bibinfo{author}{James, G.}, \bibinfo{author}{Dunaway, D.J.}, \bibinfo{year}{2017}b.
\newblock \bibinfo{title}{{Quantifying the effect of corrective surgery for trigonocephaly: A non-invasive, non-ionizing method using three-dimensional handheld scanning and statistical shape modelling}}.
\newblock \bibinfo{journal}{Journal of Cranio-Maxillo-Facial Surgery} , \bibinfo{pages}{387--394}\URLprefix \url{http://dx.doi.org/10.1016/j.jcms.2017.01.002}, \DOIprefix\doi{10.1016/j.jcms.2017.01.002}.
%Type = Article
\bibitem[{Tarnow et~al.(2022)Tarnow, K{\"{o}}lby, Maltese, S{\"{o}}fteland, Lew{\'{e}}n, Nilsson, Enblad and Nowinski}]{Tarnow2022IncidenceSweden}
\bibinfo{author}{Tarnow, P.}, \bibinfo{author}{K{\"{o}}lby, L.}, \bibinfo{author}{Maltese, G.}, \bibinfo{author}{S{\"{o}}fteland, M.B.}, \bibinfo{author}{Lew{\'{e}}n, A.}, \bibinfo{author}{Nilsson, P.}, \bibinfo{author}{Enblad, P.}, \bibinfo{author}{Nowinski, D.}, \bibinfo{year}{2022}.
\newblock \bibinfo{title}{{Incidence of Non-Syndromic and Syndromic Craniosynostosis in Sweden}}.
\newblock \bibinfo{journal}{Journal of Craniofacial Surgery} \bibinfo{volume}{33}, \bibinfo{pages}{1517--1520}.
\newblock \URLprefix \url{https://journals.lww.com/jcraniofacialsurgery/fulltext/2022/07000/incidence_of_non_syndromic_and_syndromic.57.aspx}, \DOIprefix\doi{10.1097/SCS.0000000000008457}.
%Type = Article
\bibitem[{Tenhagen et~al.(2016)Tenhagen, Bruse, Rodriguez-Florez, Angullia, Borghi, Koudstaal, Schievano, Jeelani and Dunaway}]{Tenhagen2016Three-DimensionalCraniosynostosis}
\bibinfo{author}{Tenhagen, M.}, \bibinfo{author}{Bruse, J.L.}, \bibinfo{author}{Rodriguez-Florez, N.}, \bibinfo{author}{Angullia, F.}, \bibinfo{author}{Borghi, A.}, \bibinfo{author}{Koudstaal, M.J.}, \bibinfo{author}{Schievano, S.}, \bibinfo{author}{Jeelani, O.}, \bibinfo{author}{Dunaway, D.}, \bibinfo{year}{2016}.
\newblock \bibinfo{title}{{Three-Dimensional Handheld Scanning to Quantify Head-Shape Changes in Spring-Assisted Surgery for Sagittal Craniosynostosis}}.
\newblock \bibinfo{journal}{The Journal of craniofacial surgery} \bibinfo{volume}{27}, \bibinfo{pages}{2117--2123}.
\newblock \URLprefix \url{https://journals.lww.com/jcraniofacialsurgery/fulltext/2016/11000/three_dimensional_handheld_scanning_to_quantify.43.aspx}, \DOIprefix\doi{10.1097/SCS.0000000000003108}.

\end{thebibliography}



\end{document}